\documentclass[prd,preprint,superscriptaddress,showpacs,nofootinbib,tightenlines,amsmath]{revtex4-1}
\usepackage{graphics}
\usepackage{epsfig}
\usepackage{slashed}
\usepackage{pbox, calc}
\usepackage{braket}
\usepackage{nicefrac}
\usepackage{bbm}
\begin{document}
\title{Contribution of the $a_1$ meson to the axial nucleon-to-delta transition form factors}
\author{Y. \"Unal}
\affiliation{Physics Department, \c{C}anakkale  Onsekiz Mart
University, 17100 \c{C}anakkale, Turkey}
\author{A. K\"u\c{c}\"ukarslan}
\affiliation{Physics Department, \c{C}anakkale  Onsekiz Mart
University, 17100 \c{C}anakkale, Turkey}
\author{S.~Scherer}
\affiliation{PRISMA Cluster of Excellence, Institut f\"ur
Kernphysik, Johannes Gutenberg-Universit\"at Mainz, D-55099 Mainz,
Germany}
\date{August 9, 2018}
\preprint{MITP/18-074}
\begin{abstract}
   We analyze the low-$Q^2$ behavior of the axial form factor $G_A(Q^2)$, the induced pseudoscalar form factor $G_P(Q^2)$,
and the axial nucleon-to-delta transition form factors $C^A_5(Q^2)$
and $C^A_6(Q^2)$.
   Building on the results of chiral perturbation theory, we first discuss $G_A(Q^2)$ in a chiral
effective-Lagrangian model including the $a_1$ meson and determine the relevant coupling parameters from a
fit to experimental data.
   With this information, the form factor $G_P(Q^2)$ can be predicted.
   For the determination of the transition form factor $C^A_5(Q^2)$, we make use of an SU(6) spin-flavor
quark-model relation to fix two coupling constants such that only one free parameter is left.
   Finally, the transition form factor $C^A_6(Q^2)$ can be predicted in terms of $G_P(Q^2)$,
the mean-square axial radius $\langle r^2_A\rangle$, and the
mean-square axial nucleon-to-delta  transition radius $\langle
r^2_{AN\Delta}\rangle$.
\end{abstract}
\maketitle

\section{Introduction}

   At the fundamental level, the electroweak form factors of hadrons originate
from the dynamics of the constituents of quantum chromodynamics (QCD), namely, quarks and gluons.
   While a wealth of precision data exists for the electromagnetic form factors
of the proton and, to a lesser extent, of the neutron (see, e.g., Refs.~\cite{Perdrisat:2006hj,Pacetti:2015iqa}
for a review), the nucleon form factors of the isovector axial-vector current,
the axial form factor $G_A$ and, in particular, the induced pseudoscalar
form factor $G_P$, are not as well known (see, e.g.,
Refs.~\cite{Bernard:2001rs,Gorringe:2002xx} for a review).
   A similar situation occurs in the case of the nucleon-to-delta transition form
factors.
   A considerable amount of data is available for the electromagnetic transition
form factors (see, e.g., Refs.~\cite{Tiator:2011pw,Aznauryan:2011qj} for a review),
whereas very little is known about the axial nucleon-to-delta transition form factors
\cite{Barish:1978pj,Radecky:1981fn,Kitagaki:1986ct,Kitagaki:1990vs,Androic:2012doa}.
   On the theoretical side, there have been various approaches to determining the
nucleon-to-delta transition form factors.
   Calculations have been performed in the framework of quark models \cite{Korner:1977rb,Hemmert:1994ky,Liu:1995bu,Golli:2002wy,BarquillaCano:2007yk},
chiral effective field theory \cite{Zhu:2002kh,Geng:2008bm,Procura:2008ze},
lattice QCD \cite{Alexandrou:2006mc,Alexandrou:2009vqd,Alexandrou:2013opa}, and light-cone QCD sum
rules \cite{Aliev:2007pi,Kucukarslan:2015urd}.
   Moreover, a substantial amount of work has been devoted to the question of
how to parametrize and extract the form factors from experimental data
\cite{Adler:1968tw,LlewellynSmith:1971uhs,Schreiner:1973mj,Adler:1975mt,Mukhopadhyay:1998mn,AlvarezRuso:1998hi,Sato:2003rq,Graczyk:2009qm,Hernandez:2010bx,Graczyk:2014dpa}.

   In this article, based on the results of Ref.~\cite{Schindler:2006it},
we make use of a semiphenomenological description
of the nucleon axial form factor $G_A$ and the induced pseudoscalar form
factor $G_P$ to predict, using certain model assumptions, two of the four
axial nucleon-to-delta transition form factors, namely, $C_5^A$ and $C_6^A$.
   We will assume that the exchange of the axial-vector meson $a_1(1260)$ provides
a dominant contribution to the form factor $C_5^A$ at low values of $Q^2$.
   Such a scenario was already envisaged decades ago in Ref.~\cite{Sehgal:1979tr},
where the common use of a dipole form was questioned.

\section{Axial-vector current operator in QCD}
   In terms of the up-quark and down-quark fields,
\begin{displaymath}
q(x)=\begin{pmatrix}u(x)\\d(x)\end{pmatrix},
\end{displaymath}
the Cartesian components of the isovector axial-vector current operator are defined as
\begin{equation}
A^\mu_j(x)=\bar{q}(x)\gamma^\mu\gamma_5\frac{\tau_j}{2}q(x).
\end{equation}
   In the isospin-symmetric limit, $m_u=m_d=\hat{m}$, the divergence of the isovector axial-vector current
is given by
\begin{equation}
\label{divA}
\partial_\mu A^\mu_j=i\hat m \bar q\gamma_5\tau_j q\equiv\hat m P_j,
\end{equation}
where $P_j$ is the $j$th component of the pseudoscalar quark density.
   After coupling external c-number axial-vector fields $a_{\mu j}(x)$ to the axial-vector current operators $A^\mu_j(x)$
\cite{Gasser:1983yg},
\begin{equation}
{\cal L}_{\rm ext}=\sum_{j=1}^3 a_{\mu j}(x)A^\mu_j(x),
\end{equation}
the invariant amplitude for a transition from a hadronic state $|A(p_i)\rangle$
to $|B(p_f)\rangle$, induced by a plane-wave external field of the form
$a_{\mu j}(x)=\epsilon_{\mu j}(q)e^{-iq\cdot x}$, is defined as (no summation over
$j$ implied)
\begin{equation}
\label{invariant_amplitude}
{\cal M}=i\epsilon_{\mu j}(q)\langle B(p_f)|A^\mu_j(0)|A(p_i)\rangle,
\end{equation}
where four-momentum conservation $p_f=p_i+q$ due to translational invariance is implied.

\section{Parametrization of the nucleon-to-nucleon and nucleon-to-$\Delta$ transitions}
The axial-vector current matrix element between nucleon states can be parametrized as
\cite{Schindler:2006it}
\begin{equation}
\label{parameterization_nucleon}
\langle N(p_f,s_f)|A_j^\mu(0)|N(p_i,s_i)\rangle=
\bar{u}(p_f,s_f)\left[\gamma^\mu\gamma_5 G_A(Q^2)+\frac{q^\mu}{2m_N}\gamma_5 G_P(Q^2)\right]\frac{\tau_j}{2}u(p_i,s_i),
\end{equation}
where $q=p_f-p_i$, $Q^2=-q^2$, and $m_N$ is the nucleon mass.
   The Pauli matrix $\tau_j$ has to be evaluated between nucleon isospinors.
   At $Q^2=0$, the axial form factor reduces to the axial-vector coupling constant
$g_A=1.2723\pm 0.0023$ \cite{Patrignani:2016xqp}.
   At $Q^2=m_\mu^2$, where $m_\mu$ is the muon mass, the induced pseudoscalar coupling constant is defined as
\begin{equation}
g_p=\frac{m_\mu}{2m_N}G_P(m_\mu^2).
\end{equation}
   Recently, the MuCap Collaboration obtained $g_p=8.06\pm 0.55$ \cite{Andreev:2012fj},
which is in very good agreement with the result of chiral perturbation theory
\cite{Bernard:1994wn,Fearing:1997dp}, $g_p=8.26\pm 0.23$ \cite{Bernard:2001rs}.\footnote{
   The result of older experiments has been somewhat under debate (see Table II of Ref.~\cite{Gorringe:2002xx})
with a world average of $g_p=10.5\pm 1.8$ of all ordinary muon capture experiments.}

   Introducing the spherical tensor notation \cite{Edmonds},
\begin{displaymath}
A^{\mu(1)}_{\pm 1}=\mp\frac{1}{\sqrt{2}}\left(A^\mu_1\pm i A^\mu_2\right),\quad A^{\mu(1)}_0=A^\mu_3,
\end{displaymath}
and using isospin symmetry, we express the matrix element of the spherical isospin components
($\alpha=+1,0,-1$) between a nucleon state and a $\Delta$ state as
\begin{equation}
\langle3/2,\tau_\Delta|A^{\mu(1)}_\alpha|1/2,\tau\rangle=
(1/2,\tau;1,\alpha|3/2,\tau_\Delta)\langle3/2||A^{\mu(1)}||1/2\rangle,
\end{equation}
where $\langle3/2||A^{\mu(1)}||1/2\rangle$ denotes the reduced matrix element and
$(1/2,\tau;1,\alpha|3/2,\tau_\Delta)$ is the relevant Clebsch-Gordan coefficient.
   For example, using $\langle 1/2,1/2;1,0|3/2,1/2\rangle=\sqrt{2/3}$, we obtain
the reduced matrix element in terms of the $p$ to $\Delta^+$ transition
as
\begin{equation}
\label{redphys}
\langle 3/2||A^{\mu(1)}||1/2\rangle=\sqrt{\frac{3}{2}}\langle \Delta^+|A^{\mu(1)}_0|p\rangle.
\end{equation}

   Because of its very short lifetime of the order of 10$^{-23}$ s, the $\Delta$ is not
a stable one-particle state.
   It shows up as a pole of the $S$-matrix in the complex-energy plane and a model-independent
definition of its properties should take place at a complex squared four-momentum $s=p^2=z_\Delta^2$ with a complex pole
position $z_\Delta=m_\Delta-i\Gamma_\Delta/2$ \cite{Gegelia:2009py}.
   From the experimental side, this means that one needs to look for a process which involves
the transition to an intermediate $\Delta$ comprising, as a building block, the "matrix element" one is interested in.
   For example, weak single-pion production in the $\Delta$-resonance region \cite{Adler:1968tw}, $\nu N\to {\ell} N\pi$,
contains information on both the vector and axial-vector nucleon-to-delta transitions.
   While experiments are performed for real values of the squared center-of-mass energy $s$,
an analytic continuation of the three-point function to complex values allows for an extraction of the delta properties
from the theoretical side.
   In Ref.~\cite{Gegelia:2009py}, a method applicable for spin-1/2 resonances was proposed to extract from the general vertex
only that piece surviving as the residue at the pole.
   In the vicinity of the pole, the renormalized dressed propagator of the resonance is written
as\footnote{Note that $\slashed{p}+z=\sum_{i=1}^2 w^i(p)\bar{w}^i(p)+{\cal O}(p^2-z^2)$.}
\begin{displaymath}
S(p)=\frac{1}{\slashed{p}-z}+\text{n.p.}=\frac{\slashed{p}+z}{p^2-z^2}+\text{n.p.}=
\sum_{i=1}^2 \frac{w^i(p)\bar{w}^i(p)}{p^2-z^2}+\text{n.p.},
\end{displaymath}
where n.p.~refers to nonpole, i.e.~regular terms and $w^i$, $\bar{w}^i$ are Dirac spinors with complex masses $z$.
   An external leg of the Green function is multiplied by $p^2-z^2$ and the result is then evaluated between the corresponding
Dirac spinors.
   The generalization to the case of Rarita-Schwinger vector-spinors $\bar{w}_\lambda(p,s)$  \cite{Rarita:1941mf,Kusaka} with a complex
mass $z_\Delta$ and $p^2=z_\Delta^2$ was described in Ref.~\cite{Agadjanov:2014kha}.

   Even though a description in terms of stable states does not exist, we use Dirac's bra-ket notation,
with the understanding that the relevant amplitude is extracted at the complex pole.
   The Lorentz structure of the reduced matrix element may be written as
\begin{equation}
\label{Lorentz_structure}
\langle \Delta(p_f,s_f)||A^{\mu(1)}(0)||N(p_i,s_i)\rangle
=\bar{w}_\lambda(p_f,s_f)\Gamma^{\lambda\mu}_Au(p_i,s_i).
\end{equation}
   Here, the initial nucleon is described by the Dirac spinor $u(p_i,s_i)$ with real
mass $m_N$ and $p_i^2=m_N^2$, the final $\Delta$(1232) is described via the Rarita-Schwinger
vector-spinor $\bar{w}_\lambda(p_f,s_f)$ \cite{Rarita:1941mf,Kusaka} with a complex mass $z_\Delta$ and $p_f^2=z_\Delta^2$.
   The explicit form of $\bar{w}_\lambda$ can be found in Ref.~\cite{Agadjanov:2014kha} but we will only need the
properties of Eq.~(\ref{rarita_schwinger}) below.

   In the following, it is always understood that the ``tensor'' $\Gamma^{\lambda\mu}_A$ is evaluated
between on-shell spinors $u$ and $\bar{w}_\lambda$, satisfying
\begin{align}
\label{dirac}
\slashed{p}_i u(p_i,s_i)&=m_N u(p_i,s_i),\\
\label{rarita_schwinger}
\bar{w}_\lambda(p_f,s_f)\slashed{p}_f&=z_\Delta \bar{w}_\lambda(p_f,s_f),\quad
\bar{w}_\lambda(p_f,s_f)\gamma^\lambda=0,\quad
\bar{w}_\lambda(p_f,s_f)p_f^\lambda=0.
\end{align}
   The last two equations are responsible for identifying the spin-3/2 component of the
Rarita-Schwinger vector-spinor.
   We tacitly assume that this is also true for the analytic continuation.
   The ``tensor'' $\Gamma^{\lambda\mu}_A$ contains a superposition of four Lorentz
tensors \cite{Adler:1968tw,LlewellynSmith:1971uhs}, which we choose to be \cite{Alexandrou:2009vqd,Procura:2008ze}
\begin{align}
\label{decomposition1}
\Gamma^{\lambda\mu}_A&=\frac{C_3^A(Q^2)}{m_N}\left(g^{\lambda\mu}\slashed{q}-q^\lambda\gamma^\mu\right)
+\frac{C_4^A(Q^2)}{m_N^2}\left(g^{\lambda\mu}p_f\cdot q-q^\lambda p_f^\mu\right)
+C_5^A(Q^2)g^{\lambda\mu}+\frac{C_6^A(Q^2)}{m_N^2}q^\lambda q^\mu.
\end{align}
   In particular, $C_5^A$ and $C_6^A$ correspond to the axial nucleon form factor
$G_A$ and the induced pseudoscalar form factor $G_P$, respectively.

   Equations (\ref{Lorentz_structure})--(\ref{decomposition1}) provide the general framework
for an unstable $\Delta$ resonance.
   However, in the present work, we do not calculate any loop corrections to $\Delta$-resonance properties.
   We therefore also neglect the width $\Gamma_\Delta$ such that our results for the form factors turn out
to be real.

\section{Axial-vector coupling constants in the static quark model}
   Here, we recall an SU(6) spin-flavor quark-model relation, which will be applied in
the subsequent calculations.
   In the static quark model, the operator $A_{z,3}$ is given by
\begin{displaymath}
A_{z,3}=\frac{1}{2}\sum_{i=1}^3\tau_3(i)\sigma_z(i).
\end{displaymath}
   The axial-vector coupling constant is obtained as
\begin{equation}
\label{gAdefquarkmodel}
\langle p, S_z=1/2|A_{z,3}|p, S_z=1/2\rangle=\frac{1}{2}g_A.
\end{equation}
   Inserting the appropriate quark-model wave function,
\begin{align}
\label{spin_isospin_proton_spin_up}
|p,S_z=1/2\rangle
&=\frac{1}{\sqrt{18}}\big[2(u\uparrow\,u\uparrow\,d\downarrow
+u\uparrow\,d\downarrow\,u\uparrow
+d\downarrow\,u\uparrow\,u\uparrow)\nonumber\\
&\quad
-(u\uparrow\,u\downarrow\,d\uparrow
+u\downarrow\,u\uparrow\,d\uparrow
+u\uparrow\,d\uparrow\,u\downarrow\nonumber\\
&\quad
+d\uparrow\,u\uparrow\,u\downarrow
+u\downarrow\,d\uparrow\,u\uparrow
+d\uparrow\,u\downarrow\,u\uparrow)\big],
\end{align}
one obtains
\begin{equation}
g_A=2\langle p,S_z=1/2|A_{z,3}|p,S_z=1/2\rangle
=3\langle p,S_z=1/2|\tau_3(3)\sigma_z(3)|p,S_z=1/2\rangle=\frac{5}{3}.
\end{equation}
   On the other hand, evaluating Eq.~(\ref{parameterization_nucleon}) for $\vec p_i=\vec p_f=\vec 0$
and $S_{zi}=S_{zf}=1/2$ yields
\begin{equation}
\label{gADiracspinors}
\bar{u}^{(1)}(\vec 0)\gamma^3\gamma_5 g_A u^{(1)}(\vec 0)\times\frac{1}{2}
=2m_N \frac{g_A}{2}\begin{pmatrix}1&0&0&0\end{pmatrix}\begin{pmatrix}0&0&1&0\\
0&0&0&-1\\
-1&0&0&0\\
0&1&0&0
\end{pmatrix}
\begin{pmatrix}
0&0&1&0\\
0&0&0&1\\
1&0&0&0\\
0&1&0&0
\end{pmatrix}
\begin{pmatrix}1\\0\\0\\0\end{pmatrix}
=2m_N\,\frac{g_A}{2}.
\end{equation}
   The factor $2m_N$ originates from our normalization of the Dirac spinors
(see Appendix \ref{appendix_Dirac_spinors}).
   When comparing the expression of Eq.~(\ref{gADiracspinors}) to the quark-model result of Eq.~(\ref{gAdefquarkmodel}), we have to discard this factor.

   Using
\begin{align}
\label{Delta+1/2}
|\Delta^+,S_z=1/2\rangle&=\frac{1}{3}(u\uparrow\,u\uparrow\,d\downarrow+u\uparrow\,d\downarrow\,u\uparrow
+d\downarrow\,u\uparrow\,u\uparrow\nonumber\\
&\quad+u\uparrow\,u\downarrow\,d\uparrow+u\downarrow\,u\uparrow\,d\uparrow+u\uparrow\,d\uparrow\,u\downarrow\nonumber\\
&\quad+ d\uparrow\,u\uparrow\,u\downarrow+u\downarrow\,d\uparrow\,u\uparrow+d\uparrow\,u\downarrow\,u\uparrow)
\end{align}
together with Eq.~(\ref{spin_isospin_proton_spin_up}), one obtains
for the nucleon-to-delta axial-vector transition
\begin{align}
\label{matrixelementNDeltaquarkmodel}
\langle \Delta^+,S_z=1/2|A_{z,3}|p,S_z=1/2\rangle
&=\frac{3}{2}\langle \Delta^+,S_z=1/2|\tau_3(3)\sigma_z(3)|p,S_z=1/2\rangle\nonumber\\
&=\frac{2}{3}\sqrt{2}=\frac{5}{3}\frac{2}{5}\sqrt{2}=
\frac{2}{5}\sqrt{2}g_A.
\end{align}

\section{Connection to chiral effective field theory}
   At lowest order in the quark-mass and momentum expansion, the relevant interaction Lagrangian for nucleons reads \cite{Gasser:1987rb}
\begin{equation}
{\cal L}_{\rm int}=
\frac{\texttt{g}_A}{2}\bar{\Psi}\gamma^\mu\gamma_5 u_\mu\Psi,
\end{equation}
where $\texttt{g}_A$ is the chiral limit of the axial-vector coupling constant
and
\begin{equation}
\Psi=\left(\begin{array}{c}p\\n\end{array}\right)
\end{equation}
denotes the nucleon field with two four-component Dirac fields for the proton and the neutron.
   The so-called chiral vielbein $u_\mu$ (see Chap.~4 of Ref.~\cite{Scherer:2012zzd}
for a detailed discussion) is a traceless, Hermitian, $(2\times 2)$ matrix,
\begin{displaymath}
u_\mu=i[u^\dagger(\partial_\mu-ir_\mu)u-u(\partial_\mu-il_\mu)u^\dagger]=\sum_{j=1}^3 \tau_j u_{\mu,j},
\end{displaymath}
   which involves the external fields $r_\mu=v_\mu+a_\mu$ and $l_\mu=v_\mu-l_\mu$ as well as pions.
   The latter are contained in the unimodular, unitary, $(2\times 2)$ matrix $u$:
\begin{equation}
\begin{split}
u(x)&=\textnormal{exp}\left(i\frac{\Phi(x)}{2F}\right),\\
\Phi(x)&=\sum_{j=1}^3\tau_j \phi_j(x)=
\left(\begin{array}{cc}\pi^0(x) & \sqrt{2}\pi^+(x)\\ \sqrt{2}\pi^-(x)&-\pi^0(x)\end{array}\right),\\
\end{split}
\label{eqn:pionmatrix}
\end{equation}
where $F$ denotes the pion-decay constant in the chiral limit:
$F_\pi=F\left[1+O(\hat{m})\right]=92.2$~MeV.

   The expansion of the chiral vielbein in the pion fields yields
\begin{displaymath}
u_\mu=2a_\mu-\frac{\partial_\mu\Phi}{F}+{\cal O}(v_\mu\Phi,a_\mu\Phi^2,\partial_\mu\Phi\Phi^2),
\end{displaymath}
where
\begin{displaymath}
a_\mu=\sum_{j=1}^3 \frac{\tau_j}{2} a_{\mu j}.
\end{displaymath}
   Keeping only the first term of the expansion, i.e., the replacement $u_\mu\to 2 a_\mu$, gives rise to the
interaction Lagrangian\footnote{On the other hand, the second term results in the pseudovector pion-nucleon
interaction,
\begin{displaymath}
{\cal L}_{\pi NN}=-\frac{\texttt{g}_A}{2F}\bar{\Psi}\gamma^\mu\gamma_5 \partial_\mu\Phi\Psi.
\end{displaymath}
}
\begin{equation}
\label{LaNN}
{\cal L}_{\rm int}= \sum_{j=1}^3a_{\mu j}
\frac{\texttt{g}_A}{2}\bar{\Psi}\gamma^\mu\gamma_5 \tau_j\Psi.
\end{equation}
   The invariant amplitude for $a_{\mu j}(x)=\epsilon_{\mu j}(q) e^{-iq\cdot x}$, with $j$ fixed, reads
\begin{displaymath}
{\cal M}=i\epsilon_{\mu j}(q) \texttt{g}_A \bar{u}(p_f)\gamma^\mu\gamma_5 \frac{\tau_j}{2}u(p_i).
\end{displaymath}
   A comparison with Eqs.~(\ref{invariant_amplitude}) and (\ref{parameterization_nucleon}) yields
\begin{equation}
\label{GAlo}
G_A(Q^2)=\texttt{g}_A.
\end{equation}
   At lowest order, there is no $Q^2$ dependence and $G_A(Q^2)$ reduces
to the axial-vector coupling constant in the chiral limit.

   For the nucleon-to-delta transition the lowest-order Lagrangian is given by [see Eq.~(4.200) of Ref.~\cite{Scherer:2012zzd} with $\tilde z=-1$]
\begin{align}
{\cal L}^{(1)}_{\pi N\Delta}&=\texttt{g}\sum_{i,j=1}^3\bar{\Psi}_{\lambda,i}\xi^\frac{3}{2}_{ij}(g^{\lambda\mu}-\gamma^\lambda\gamma^\mu)
u_{\mu,j}\Psi+\text{H.c.}\nonumber\\
&\to\texttt{g}\sum_{i,j=1}^3\bar{\Psi}_{\lambda,i}\xi^\frac{3}{2}_{ij}(g^{\lambda\mu}-\gamma^\lambda\gamma^\mu)a_{\mu j}\Psi+\text{H.c.},
\label{LaND}
\end{align}
where $\Psi_{\lambda,i}$ denotes a vector-spinor isovector-isospinor field.
   The isovector-isospinor transforms under the $1\otimes\frac{1}{2}=\frac{3}{2}\oplus\frac{1}{2}$ representation
and, thus, contains both isospin 3/2 and isospin 1/2 components.
   In order to describe the $\Delta$, it is necessary to project onto the isospin-3/2 subspace.
   The corresponding matrix representation of the projection operator is denoted by $\xi^\frac{3}{2}$ and
the entries are given by \cite{Scherer:2012zzd}
\begin{displaymath}
\xi^\frac{3}{2}_{ij}=\delta_{ij}-\frac{1}{3}\tau_i\tau_j.
\end{displaymath}
   Furthermore, in order to identify the coupling to the external axial-vector field $a_{\mu j}$, we made, as in
the nucleon case, the replacement $u_{\mu,j}\to a_{\mu j}$.
   Considering $j=3$ and making use of Eq.~(4.184) of Ref.~\cite{Scherer:2012zzd},
\begin{displaymath}
\bar{\Psi}_{\lambda,i}\xi^{\frac{3}{2}}_{i3}=
\sqrt{\frac{2}{3}}\begin{pmatrix}\bar{\Delta}^+_\lambda&\bar{\Delta}^0_\lambda\end{pmatrix},
\end{displaymath}
we obtain
\begin{displaymath}
{\cal L}_{\rm int}=\sqrt{\frac{2}{3}}\texttt{g}\begin{pmatrix}\bar{\Delta}^+_\lambda&\bar{\Delta}^0_\lambda\end{pmatrix}
(g^{\lambda\mu}-\gamma^\lambda\gamma^\mu)a_{\mu3}\begin{pmatrix}p\\n\end{pmatrix}+\text{H.c.}
\end{displaymath}
   Using Eq.~(\ref{rarita_schwinger}), the invariant amplitude of $p\to\Delta^+$ reads
\begin{displaymath}
{\cal M}=i\sqrt{\frac{2}{3}}\texttt{g}\bar{w}_\lambda(p_f,s_f)g^{\lambda\mu}u(p_i,s_i)\epsilon_{\mu 3}(q).
\end{displaymath}
   The reduced matrix element [see Eq.~(\ref{redphys})] is obtained by multiplying by
$\sqrt{3/2}$ and crossing out the factors $i$ and $\epsilon_{\mu 3}(q)$:
\begin{equation}
\bar{w}_\lambda(p_f,s_f)\Gamma^{\lambda\mu}_A u(p_i,s_i)=\texttt{g}\bar{w}_\lambda(p_f,s_f)g^{\lambda\mu}u(p_i,s_i).
\end{equation}
   A comparison with Eq.~(\ref{decomposition1}) then yields the analogue of Eq.~(\ref{GAlo}), namely,
\begin{equation}
\label{CA5lo}
C_5^A(Q^2)=\texttt{g}.
\end{equation}
   Finally, how is this related to the static quark model? For this purpose, we consider
\begin{align*}
\langle\Delta^+(\vec 0),S_z=1/2|A_{z,3}(0)|p(\vec 0),S_z=1/2\rangle
&=-\sqrt{\frac{2}{3}}\texttt{g}\bar{w}_3(\vec 0,S_z=1/2)u(\vec 0,S_z=1/2),
\end{align*}
where we made use of $g^{33}=-1$.
   Since $\epsilon_{3,3}=-1$ and $\epsilon_{1,3}+i\epsilon_{2,3}=0$, we obtain in terms of the appropriate
Clebsch-Gordan coefficients,
\begin{align*}
w_3(\vec 0,S_z=1/2)&=\sqrt{\frac{2}{3}}\epsilon_{3,3}u(\vec 0, S_z=1/2)
+\frac{1}{\sqrt{3}}\left(-\frac{1}{\sqrt{2}}(\epsilon_{1,3}+i\epsilon_{2,3})\right)
u(\vec 0,S_z=-1/2)\\
&=-\sqrt{\frac{2}{3}}u(\vec 0,S_z=1/2).
\end{align*}
   Putting the pieces together, the matrix element is given by
\begin{align}
\langle\Delta^+(\vec 0),S_z=1/2|A_{z,3}(0)|p(\vec 0),S_z=1/2\rangle
&=-\sqrt{\frac{2}{3}}\texttt{g}(-1)\sqrt{\frac{2}{3}}\bar{u}(\vec 0,S_z=1/2)u(\vec 0,S_z=1/2)\nonumber\\
&=\texttt{g}\frac{2}{3}\sqrt{2m_\Delta}\sqrt{2m_N}.
\end{align}
   Again, when we compare this to Eq.~(\ref{matrixelementNDeltaquarkmodel}) for the static quark model, we have to cross out the normalization factors
$\sqrt{2m_\Delta}$ and $\sqrt{2m_N}$.
   In combination with Eq.~(\ref{GAlo}) we obtain
\begin{displaymath}
\frac{2}{3}\texttt{g}=\frac{2}{5}\sqrt{2}\,\texttt{g}_A,
\end{displaymath}
or
\begin{equation}
\texttt{g}=\frac{3}{5}\sqrt{2}\,\texttt{g}_A.
\label{relggA}
\end{equation}
   In Table \ref{table_parameters}, we collect the numerical values of the masses and coupling constants which are
taken as fixed in the subsequent calculations.
\renewcommand{\arraystretch}{1.3}
\begin{table}[h]
\caption{Masses and coupling constants.
\label{table_parameters}}
\begin{center}
\begin{tabular}{l l}
\hline
\hline
Pion mass \quad&\quad $M_\pi=139.57$~MeV\\
Nucleon mass \quad&\quad $m_N=938.92$~MeV\\
$a_1$ mass \quad&\quad $M_{a_1}=1260$~MeV\\
Pion-decay constant \quad&\quad $F_\pi=92.2$~MeV\\
Axial-vector coupling constant \quad&\quad $g_A=1.2723$\\
Pion-nucleon coupling constant \quad&\quad $g_{\pi N}^2/(4\pi)=13.69$\\
\hline
\hline
\end{tabular}
\end{center}
\end{table}
\renewcommand{\arraystretch}{1}

\section{Inclusion of the $a_1$ axial-vector meson}
   The vector mesons $\rho$ and $\omega$ play an important role in the description
of the electromagnetic form factors of the nucleon in chiral effective field theory
\cite{Kubis:2000zd,Schindler:2005ke,Bauer:2012pv}.
   Similarly, the $a_1$ axial-vector meson leads to an improved description of
the axial form factor $G_A$ \cite{Schindler:2006it}.
   Moreover, in the $\gamma^\ast N\to\Delta$ transition, the contribution
of the $\rho$ meson is needed to obtain a good description of the experimental
data \cite{Hilt:2017iup}.
   Even though we have rather little experimental data for the axial $N\Delta$ transition
   \cite{Barish:1978pj,Radecky:1981fn,Kitagaki:1986ct,Kitagaki:1990vs,Androic:2012doa},
we expect that the $a_1$ meson plays a similar role as in the
nucleon case.
   For that reason, we discuss the relevant Lagrangians and calculate
their contribution to the form factors.

\subsection{Nucleon}
   The Lagrangian for the interaction of the $a_1$ meson with the building block
$f_{-\mu\nu}$ is given by [see Eq.~(52) of Ref.~\cite{Ecker:1989yg}]
\begin{equation}
-\frac{1}{4}f_A\langle A^{\mu\nu}f_{-\mu\nu}\rangle,
\end{equation}
where $\langle\ldots\rangle$ denotes $\text{Tr}(\ldots)$ and
\begin{align*}
A^{\mu\nu}&=\nabla^\mu A^\nu-\nabla^\nu A^\mu,\\
\nabla^\mu A^\nu&=\partial^\mu A^\nu+[\Gamma^\mu,A^\nu],\\
\Gamma^\mu&=\frac{1}{2}\left[u^\dagger(\partial^\mu-ir^\mu)u+u(\partial^\mu-il^\mu)u^\dagger\right],\\
f_{-\mu\nu}&=uf_{L\mu\nu}u^\dagger-u^\dagger f_{R\mu\nu}u,\\
f_{L\mu\nu}&=\partial_\mu l_\nu-\partial_\nu l_\mu-i[l_\mu,l_\nu],\\
f_{R\mu\nu}&=\partial_\mu r_\nu-\partial_\nu r_\mu-i[r_\mu,r_\nu].
\end{align*}
   In comparison with Ref.~\cite{Ecker:1989yg}, we omit the roof sign; i.e., we write $A^\mu$ instead of $\hat A^\mu$.
Moreover, we introduce an additional factor $1/\sqrt{2}$, because our normalization of the field matrix is
\begin{displaymath}
A_\mu=\sum_{i=1}^3A_{\mu i} \tau_i,
\end{displaymath}
whereas Ecker {\it et al.} use [see Eq.~(3.4) of Ref.~\cite{Ecker:1988te}]
\begin{displaymath}
A_\mu=\frac{1}{\sqrt{2}}\sum_{i=1}^3A_{\mu i} \tau_i .
\end{displaymath}
   The replacement
\begin{align*}
r^\mu&\to a^\mu,\quad
l^\mu\to-a^\mu,\quad
f^{\mu\nu}_-\to-2(\partial^\mu a^\nu-\partial^\nu a^\mu),
\end{align*}
results in the interaction Lagrangian
\begin{align*}
\frac{1}{2}f_A\langle A^{\mu\nu}(\partial_\mu a_\nu-\partial_\nu a_\mu)\rangle=\frac{f_A}{2}
(\partial^\mu A^\nu_i-\partial^\nu A^\mu_i)(\partial_\mu a_{\nu i}-\partial_\nu a_{\mu i}).
\end{align*}
   The invariant amplitude for the coupling of an incoming external axial source
with four-momentum $q$, polarization vector $\epsilon$, and isospin component $3$
to an outgoing $a_1$ meson with four-momentum $q$, polarization vector $\epsilon_A$, and
isospin component $3$ reads
\begin{equation}
\label{matoa1}
{\cal M}=i\frac{f_A}{2}(iq^\mu\epsilon_A^{\nu\ast}-iq^\nu\epsilon_A^{\mu\ast})
(-iq_\mu\epsilon_\nu+iq_\nu\epsilon_\mu)
=if_A \epsilon_{A\nu}^\ast(q^2g^{\nu\mu}-q^\nu q^\mu)\epsilon_\mu.
\end{equation}

   The lowest-order Lagrangian for the interaction of the $a_1$ meson with the nucleon is
given by [see Eq.~(20) of Ref.~\cite{Schindler:2006it}]
\begin{equation}
{\cal L}_{a_1N}=\frac{g_{a_1N}}{2}\bar{\Psi}\gamma^\mu\gamma_5 A_\mu\Psi.
\end{equation}
   The corresponding Feynman rule for the absorption of an $a_1$ meson with
isospin index $i$ reads
\begin{displaymath}
i\frac{g_{a_1N}}{2}\gamma^\mu\gamma_5 \tau_i.
\end{displaymath}

\begin{figure}[h]
\begin{center}
\includegraphics[width=0.4\textwidth]{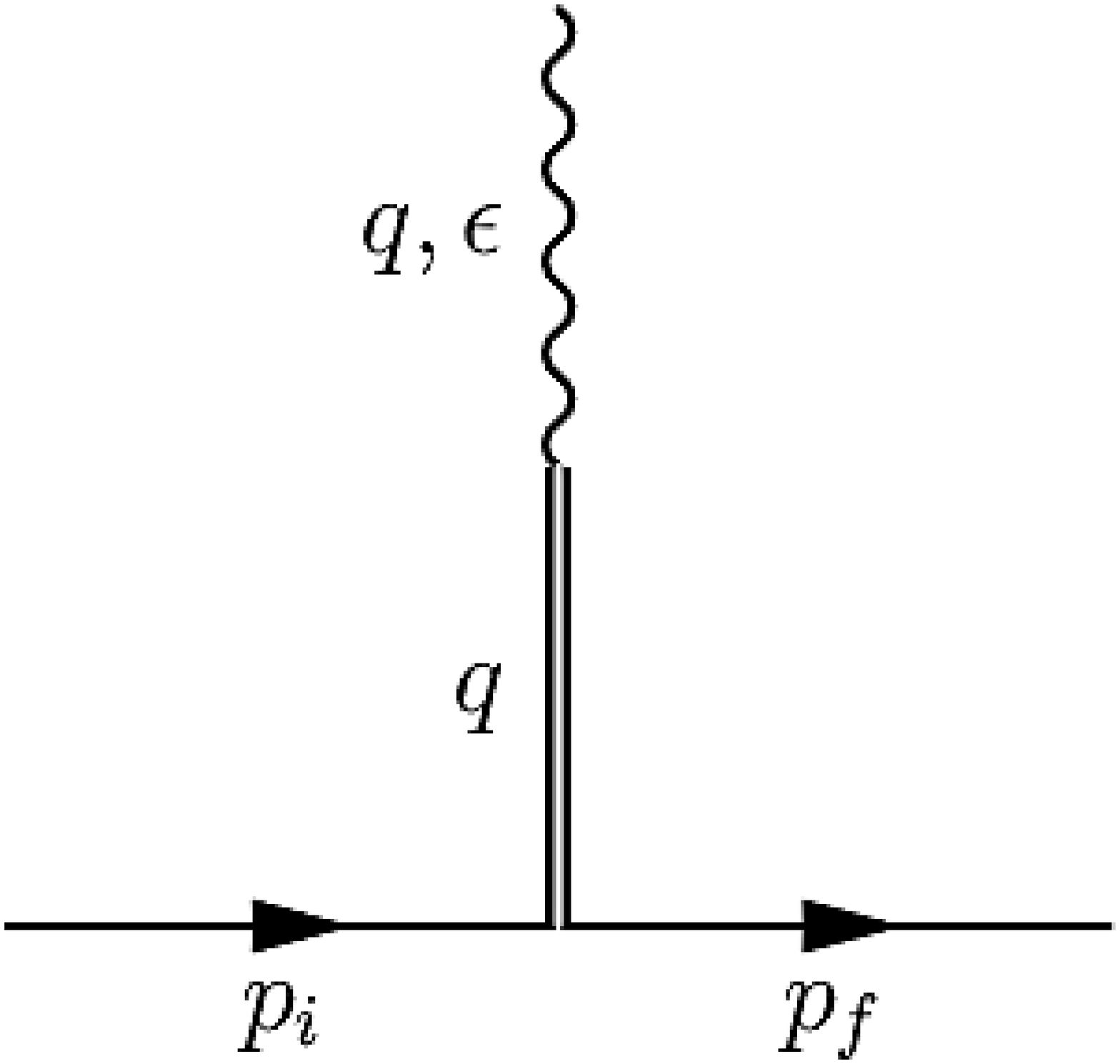}
\caption{\label{figure_a1_contribution} $a_1$ contribution to the axial-vector current matrix
element between nucleon states.}
\end{center}
\end{figure}
  The contribution to the invariant amplitude for the axial-vector transition induced
by $a_{\mu j}(x)=\epsilon_{\mu j}(q)e^{-iq\cdot x}$ is then given by (see Fig.~\ref{figure_a1_contribution})
\begin{align*}
{\cal M}=i\frac{g_{a_1N}}{2}\bar{u}(p_f,s_f)\gamma^\rho\gamma_5\tau_j u(p_i,s_i)
\left(-g_{\rho\nu}+\frac{q_\rho q_\nu}{M^2_{a_1}}\right)\frac{i}{q^2-M_{a_1}^2}
if_A(q^2 g^{\nu\mu}-q^\nu q^\mu)\epsilon_{\mu j}(q).
\end{align*}
   Note that
\begin{displaymath}
q_\nu(q^2 g^{\nu\mu}-q^\nu q^\mu)=q^\mu q^2 -q^2 q^\mu=0.
\end{displaymath}
   We thus obtain
\begin{align*}
{\cal M}=i\frac{f_Ag_{a_1N}}{2}\frac{1}{q^2-M_{a_1}^2}
\left[q^2\bar u(p_f,s_f)\gamma^\mu\gamma_5\tau_j u(p_i,s_i)-q^\mu\bar{u}(p_f,s_f)\slashed{q}\gamma_5\tau_j u(p_i,s_i)\right]\epsilon_{\mu j}(q).
\end{align*}
   Making use of $\bar u(p_f,s_f)\slashed q\gamma_5 u(p_i,s_i)=2m_N\bar u(p_f,s_f)\gamma_5 u(p_i,s_i)$,
we can then read off the contributions to $G_A$ and $G_P$:\footnote{
Note that, due to a typo, Eqs.~(46) and (47) of
Ref.~\cite{Schindler:2006it} contain an overall opposite sign.}
\begin{align}
\label{contrGA}
G_A:&\quad f_A g_{a_1N} \frac{Q^2}{M_{a_1}^2+Q^2},\\
\label{contrGP}
G_P:&\quad f_Ag_{a_1N}\frac{4m_N^2}{M_{a_1}^2+Q^2}.
\end{align}

   In essence, the loop diagrams play no role in the one-loop calculation of the
axial form factor $G_A$.
   In terms of the low-energy constants (LECs) of the Lagrangian of Ref.~\cite{Fettes:2000gb}, one obtains
\begin{equation}
\label{GAChPT}
G_A(Q^2)=\texttt{g}_A+4d_{16}M_\pi^2-d_{22}Q^2,
\end{equation}
where $d_{16}$ provides a quark-mass correction to the axial-vector coupling constant,
$g_A=\texttt{g}_A+4d_{16}M_\pi^2$, and $d_{22}$ is related to the mean-square axial radius.
   In other words, the low-$Q^2$ behavior is encoded in two constants $g_A$ and $\langle r^2_A\rangle$ which
chiral symmetry does not predict:
\begin{equation}
\label{Galinear}
G_A^{\rm linear}(Q^2)=g_A\left(1-\frac{1}{6}\langle r_A^2\rangle Q^2\right).
\end{equation}
    Experimental data are commonly analyzed in terms of the dipole parametrization,
\begin{equation}
\label{GAdipole}
G_A^{\rm dipole}(Q^2)=\frac{g_A}{\left(1+\frac{Q^2}{M_A^2}\right)^2},
\end{equation}
where the parameter $M_A$ is referred to as the axial mass.
   The weighted average extracted from (quasi)elastic neutrino and antineutrino
scattering experiments is $M_A=(1.026\pm 0.021)$~GeV \cite{Bernard:2001rs} corresponding to
a mean-square axial radius $\langle r_A^2\rangle=(0.444\pm 0.018)$~fm$^2$.
   A subsequent re-analysis of quasielastic data on deuterium has reported
$M_A=(1.016\pm0.026)$~GeV [$\langle r_A^2\rangle=(0.453\pm0.023)$~fm$^2$] \cite{Bodek:2007ym}.
   Table \ref{table_MA_exp} shows the results for $M_A$ reported by more recent experiments
on neutrino-nucleus cross sections.
   The extraction of the single-nucleon form factors from data on nuclei is a challenging
endeavor.
   Therefore, these numbers have to be treated with some caution, because they
heavily rely on the theoretical input/model used in the extraction.
   In particular, some important ingredients were previously missing such as the $n$ particle $n$ hole excitation mechanism
proposed in Ref.~\cite{Martini:2009uj}.
   Moreover, in contrast to electron-scattering experiments, the extraction is made more
complex by the fact that one has to deal with a spectrum of incident neutrinos rather than a monochromatic neutrino beam.
   For a detailed review on both the experimental and theoretical sides of this topic, see Ref.~\cite{Katori:2016yel}.
   For a discussion of theoretical studies abandoning the dipole form in their analyses, see, e.g.,
Refs.~\cite{Bhattacharya:2011ah,Bhattacharya:2015mpa,Meyer:2016oeg,Hill:2017wgb,Alvarez-Ruso:2018rdx}.
   The weighted average extracted from charged pion electroproduction experiments
is $M_A=(1.069\pm 0.016)$~GeV \cite{Bernard:2001rs} resulting in
$\langle r_A^2\rangle=(0.409\pm0.012)$~fm$^2$.

\renewcommand{\arraystretch}{1.3}
\begin{table}[h]
\caption{Axial masses reported by recent
(quasi)elastic neutrino and antineutrino scattering experiments.
\label{table_MA_exp}}
\begin{center}
\begin{tabular}{l c}
\hline
\hline
Experiment &\quad $M_A$ [GeV] \\
\hline
K2K \cite{Gran:2006jn} &\quad $1.20 \pm 0.12$\\
NOMAD \cite{Lyubushkin:2008pe} &\quad $1.05\pm 0.06$\\
MiniBooNE \cite{AguilarArevalo:2010zc} &\quad $1.35\pm 0.17$\\
MINERvA \cite{Fields:2013zhk} &\quad $0.99$\\
MINOS \cite{Adamson:2014pgc} &\quad $1.23^{+0.13}_{-0.09}(\text{fit})^{+0.12}_{-0.15}(\text{syst})$\\
\hline
\hline
\end{tabular}
\end{center}
\end{table}
\renewcommand{\arraystretch}{1}

   Including the $a_1$ meson, the axial form factor may be written as
\begin{align}
\label{GAfulla}
G_A(Q^2)&=g_A+c_1 Q^2+c_2 \frac{Q^2}{M_{a_1}^2+Q^2}\\
\label{GAfull}
&=g_A\left[1+\tilde{c}_1 Q^2-\tilde{c}_2\frac{(Q^2)^2}{M_{a_1}^2(M^2_{a_1}+Q^2)}\right],
\end{align}
where $g_A \tilde{c}_1=c_1+c_2/M^2_{a_1}$ and $g_A \tilde{c}_2=c_2=f_A g_{a_1N}$.
   The structure of the first two terms on the right-hand side of Eq.~(\ref{GAfulla})
is the same as that of Eq.~(\ref{GAChPT}) but one has to keep in mind that the LEC
$d_{22}$ will have a different value in the theory including the $a_1$ meson.
   Introducing the normalized axial form factor as
\begin{equation}
F_A(Q^2)=\frac{G_A(Q^2)}{G_A(0)},
\end{equation}
   the parametrization of $F_A(Q^2)$ contains two parameters, namely, $\tilde{c}_1$
and $\tilde{c}_2$, which can be determined from a fit to experimental data.
   Expanding the normalized axial form factor as
\begin{equation}
F_A(Q^2)=1-\frac{1}{6}\langle r_A^2\rangle Q^2+\frac{1}{120}\langle r_A^4\rangle (Q^2)^2+\cdots,
\end{equation}
for the parametrization including the $a_1$ meson, Eq.~(\ref{GAfull}),
we can identify the mean-square and mean-quartic axial radii as
\begin{equation}
\label{rA2a1}
\langle r_A^2\rangle=-6\tilde{c}_1,\quad \langle r_A^4\rangle=-120 \frac{\tilde{c}_2}{M_{a_1}^4},
\end{equation}
respectively.
   On the other hand, for the dipole parametrization, Eq.~(\ref{GAdipole}),
one obtains
\begin{equation}
\label{rA2dip}
\langle r_A^2\rangle=\frac{12}{M_A^2},\quad \langle r_A^4\rangle=\frac{360}{M_A^4}.
\end{equation}

\begin{figure}[h]
\begin{center}
\includegraphics[width=0.8\textwidth]{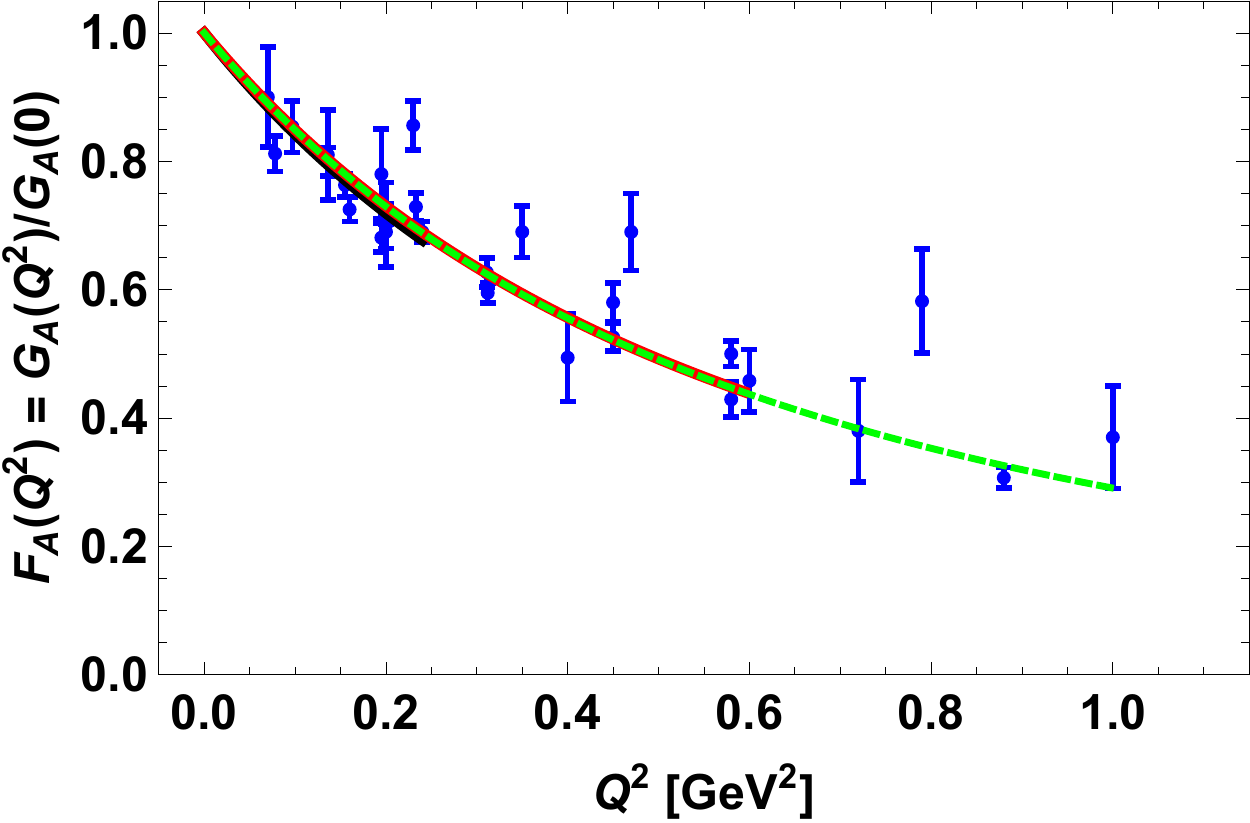}
\caption{\label{figure_dipole_fit} (Color online) $F_A(Q^2)=G_A(Q^2)/G_A(0)$ fitted to different ranges of momentum transfer $Q^2$
using the dipole parametrization of Eq.~(\ref{GAdipole}).
The (black) solid line corresponds to a fit up to and including $Q^2_{\rm max}=0.24$ GeV$^2$, the (red) long-dashed line
up to and including $Q^2_{\rm max}=0.6$ GeV$^2$, and the (green) short-dashed line up to and including $Q^2_{\rm max}=1$ GeV$^2$,
respectively.
The corresponding parameters are given in Table \ref{radii_dipole}.}
\end{center}
\end{figure}

   Figure \ref{figure_dipole_fit} shows the results of fitting the dipole parametrization to experimental data
extracted from pion electroproduction experiments \cite{Bernard:2001rs}.\footnote{We would like to
thank U.-G.~Mei{\ss}ner for providing the data in the form of a table.}
   The fits are performed for different values of the maximal squared momentum
transfer, $Q^2_{\rm max}$, and the corresponding axial masses, mean-square axial radii, and mean-quartic axial radii are
summarized in Table \ref{radii_dipole}.\footnote{Strictly
speaking, because of a loop correction to the threshold electric dipole amplitude $E_{0+}$,
the mean-square axial radius extracted from pion electroproduction has to be modified by an amount
\begin{displaymath}
\frac{3}{64 F_\pi^2}\left(\frac{12}{\pi^2}-1\right)=0.0456\,\text{fm}^2,
\end{displaymath}
such that the true axial radius is slightly larger \cite{Bernard:2001rs,Bernard:1992ys}.
   This is consistent with the observation that the average for $M_A$ extracted from charged pion
electroproduction experiments is larger than the value from (quasi)elastic neutrino and
antineutrino scattering experiments.}
   In their common domain, the curves associated with $Q^2_{\rm max}=0.6$~GeV$^2$ and $Q^2_{\rm max}=1$~ GeV$^2$
are hardly distinguishable in Fig.~\ref{figure_dipole_fit}, because the difference between the fitted axial masses is very small.

\renewcommand{\arraystretch}{1.3}
\begin{table}[h]
\caption{Comparison of the axial masses, mean-square axial radii, and mean-quartic axial radii obtained
from the dipole expression of the form factor $F_A$ fitted to different ranges of
momentum transfer.
\label{radii_dipole}}
\begin{center}
\begin{tabular}{l c c c c}
\hline
\hline
$Q^2_{\rm max}$ [GeV$^2$] &\quad $M_A$ [GeV] &\quad $\langle r^2_A\rangle$ [fm$^2$] &
\quad $\langle r_A^4\rangle$ [fm$^4$] &\quad $\chi^2_{\rm red}$\\
\hline
$0.24$ &\quad $1.057 \pm 0.027$ &\quad $0.418 \pm 0.021$ &\quad $0.437 \pm 0.045$ &\quad 2.87\\
$0.6$ &\quad $1.084 \pm 0.020$ &\quad $0.398 \pm 0.015$ &\quad $0.395 \pm 0.029$ &\quad 3.21\\
$1.0$ &\quad $1.082 \pm 0.019$ &\quad $0.399 \pm 0.014$ &\quad $0.398 \pm 0.028$ &\quad 2.97\\
\hline
\hline
\end{tabular}
\end{center}
\end{table}
\renewcommand{\arraystretch}{1}

\begin{figure}[h]
\begin{center}
\includegraphics[width=0.8\textwidth]{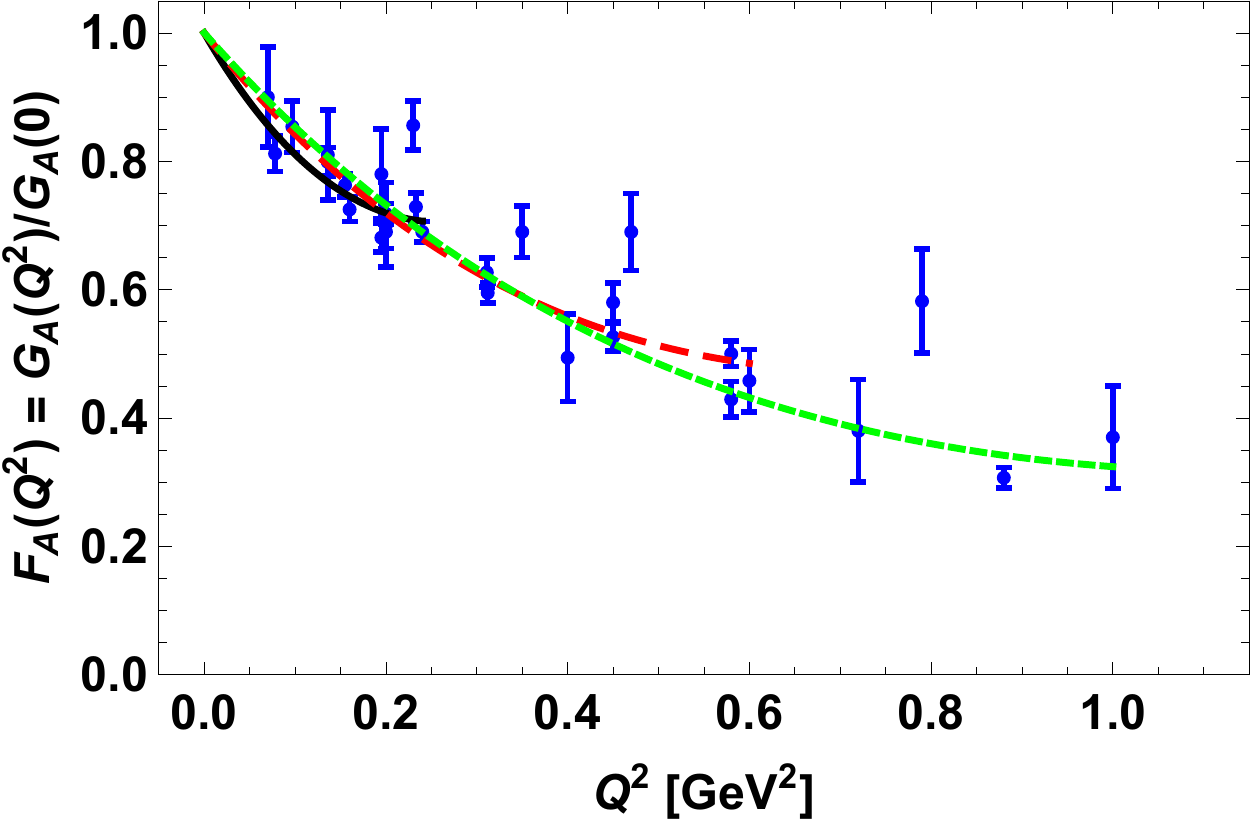}
\caption{\label{figure_a1_fit} (Color online) $F_A(Q^2)=G_A(Q^2)/G_A(0)$ fitted to different ranges of momentum transfer $Q^2$
using the parametrization of Eq.~(\ref{GAfull}) including the $a_1$ meson.
The (black) solid line corresponds to a fit up to and including $Q^2_{\rm max}=0.24$ GeV$^2$, the long-dashed (red) line
up to and including $Q^2_{\rm max}=0.6$ GeV$^2$, and the short-dashed (green) line up to and including $Q^2_{\rm max}=1$ GeV$^2$,
respectively.
The corresponding parameters are given in Table \ref{radii_a1}.}
\end{center}
\end{figure}

   Figure \ref{figure_a1_fit} shows the corresponding fits using the parametrization of Eq.~(\ref{GAfull}) including the
$a_1$ meson ($a_1$ fits for short).
   The respective parameters $\tilde{c}_1$ and $\tilde{c}_2$, mean-square axial radii, and mean-quartic axial radii are summarized in
Table \ref{radii_a1}.
   When comparing the $a_1$ fit to the dipole fit, one should keep in mind that Eq.~(\ref{GAfull}) represents a model for the low-$Q^2$
behavior of the axial form factor with a restricted domain of validity.
   The fits of Fig.~\ref{figure_a1_fit} share the common feature that $F_A$, when extrapolated beyond $Q^2_{\rm max}$,
very soon starts to rise again and diverges as $Q^2\to\infty$.
   This is, of course, an unphysical feature, originating from the linear term proportional to $c_1$ in Eq.~(\ref{GAfulla}).
   Moreover, the $a_1$ contribution asymptotically does not fall off as $1/(Q^2)^2$ as predicted by perturbative QCD
\cite{Carlson:1985zu}.\footnote{Note that the dipole form shows this
behavior.}
   Motivated by the observation that the dipole fit and the $a_1$ fit produce similar results for $Q^2_{\rm max}=0.6$~GeV$^2$,
we will, somewhat arbitrarily, assume that this value provides a
reasonable upper limit for the range of applicability of the $a_1$
model.
   According to Eqs.~(\ref{GAfulla}) and (\ref{rA2a1}), the mean-square axial radius obtains a contribution from both
the low-energy constant (LEC) $c_1$ and the $a_1$-pole diagram (see Fig.~\ref{figure_a1_contribution}).
   For the values of $\tilde{c}_1$ and $\tilde{c}_2$ of Table \ref{radii_a1}, the $a_1$ contribution to $\langle r_A^2\rangle$
is larger than the total result, implying a negative contribution from the LEC $c_1$.
   To be specific, for $Q^2_{\rm max}=0.6$~GeV$^2$ we obtain $\langle r_A^2\rangle_{\rm LEC}+\langle r_A^2\rangle_{a_1}=
(-0.366+0.781)\,\text{fm}^2=0.415\,\text{fm}^2$.

\renewcommand{\arraystretch}{1.3}
\begin{table}[h]
\caption{Comparison of the parameters $\tilde{c}_1$ and $\tilde{c}_2$, mean-square axial radii, and mean-quartic axial radii obtained
from the expression Eq.~(\ref{GAfulla}) of the form factor $F_A$ fitted to different ranges of
momentum transfer.
\label{radii_a1}}
\begin{center}
\begin{tabular}{l c c c c c}
\hline
\hline
$Q^2_{\rm max}$ [GeV$^2$] &\quad $\tilde{c}_1$ [GeV$^{-2}$] &\quad$\tilde{c}_2$ &\quad $\langle r_A^2\rangle$ [fm$^2$]
&\quad $\langle r_A^4\rangle$ [fm$^4$] &\quad $\chi^2_{\rm red}$\\
\hline
$0.24$ &\quad $-2.44 \pm 0.32$ &\quad $-14.8 \pm 4.4$ &\quad $0.570 \pm 0.075$ &\quad $1.068 \pm 0.318$ &\quad 2.08\\
$0.6$ &\quad $-1.78 \pm 0.09$ &\quad $-5.31 \pm 0.69$ &\quad $0.416 \pm 0.021$ &\quad $0.383 \pm 0.050$ &\quad 2.68\\
$1.0$ &\quad $-1.61 \pm 0.07$ &\quad $-3.84 \pm 0.39$ &\quad $0.376 \pm 0.016$ &\quad $0.277 \pm 0.028$ &\quad 3.27\\
\hline
\hline
\end{tabular}
\end{center}
\end{table}
\renewcommand{\arraystretch}{1}

\begin{figure}[h]
\begin{center}
\includegraphics[width=0.8\textwidth]{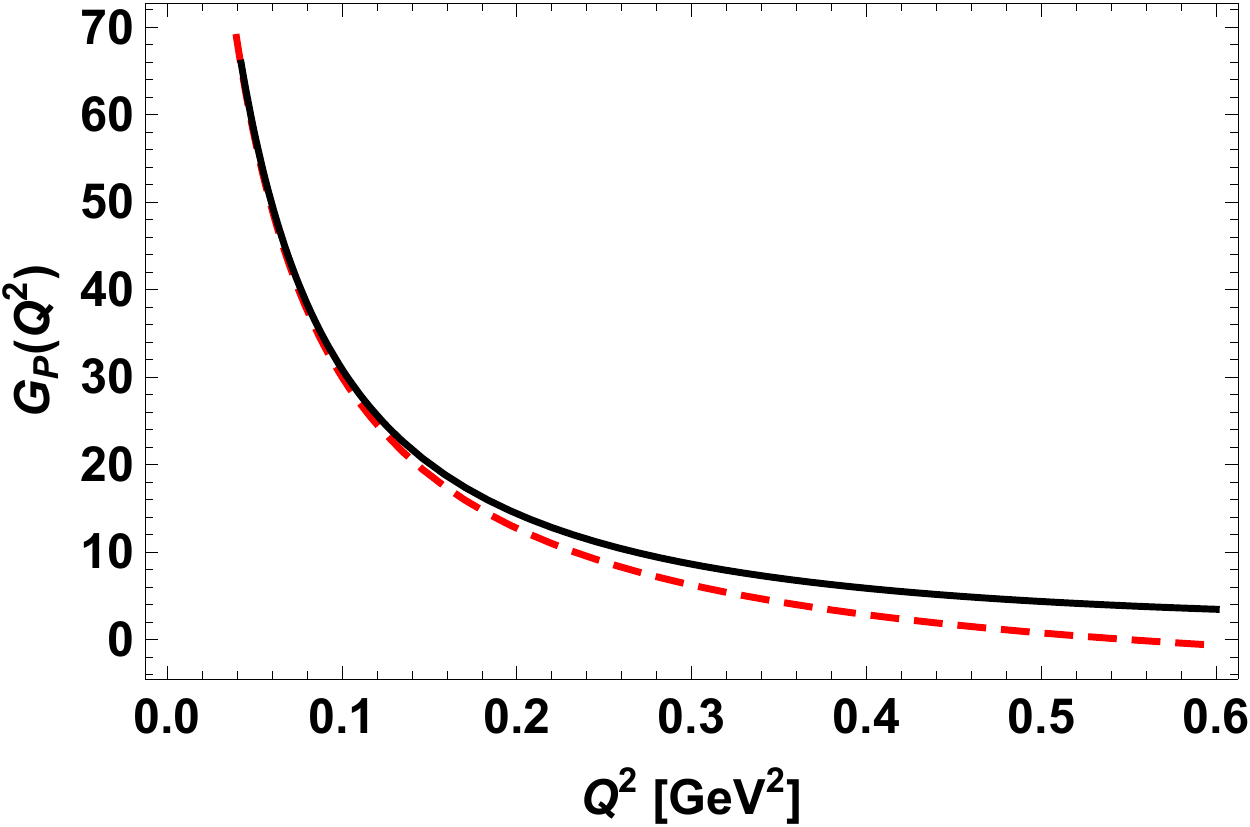}
\caption{\label{figure_GP} (Color online) Induced pseudoscalar form factor $G_P(Q^2)$.
(Red) dashed line: Pion-pole-dominance result. (Black) solid line: Result including
the $a_1$ meson.}
\end{center}
\end{figure}

   At order ${\cal O}(p^3)$ in chiral perturbation theory, the low-$Q^2$ behavior of the induced pseudoscalar form factor $G_P(Q^2)$
can entirely be written in terms of known physical quantities \cite{Schindler:2006it,Bernard:1994wn},
\begin{equation}
\label{GPexp}
G_P(Q^2)=4\frac{m_N F_\pi g_{\pi N}}{M_\pi^2+Q^2}-\frac{2}{3}m_N^2 g_A\langle r_A^2\rangle,
\end{equation}
where $g_{\pi N}$ denotes the pion-nucleon coupling constant with
$g_{\pi N}^2/(4\pi)=13.69\pm 0.19$ \cite{Baru:2011bw}.
   Using Eq.~(\ref{contrGP}), the relevant expression including the $a_1$ meson reads
\begin{equation}
\label{GPa1}
G_P(Q^2)=4\frac{m_N F_\pi g_{\pi N}}{M_\pi^2+Q^2}-\frac{2}{3}m_N^2 g_A\langle r_A^2\rangle
-4m_N^2 g_A\tilde{c}_2 \frac{Q^2}{M_{a_1}^2(M_{a_1}^2+Q^2)},
\end{equation}
where the mean-square axial radius is given in Eq.~(\ref{rA2a1}).
   In Fig.~\ref{figure_GP}, we compare the results for $G_P(Q^2)$ including the $a_1$ contribution (solid line)
and without the $a_1$ contribution (dashed line).
   Clearly, at low $Q^2$, the form factor is dominated by the pion-pole contribution and a
deviation due to the $a_1$ meson is only seen for larger values of $Q^2$, where the form factor
is small.

\subsection{Nucleon-to-delta transition}
   In order to discuss the $a_1$-meson contribution to $C_5^A(Q^2)$ and $C_6^A(Q^2)$, we need the coupling of
the $a_1$ meson to the $N\Delta$ system.
   We model this interaction in analogy to the coupling of the external axial-vector field $a_{\mu j}(x)$
[see Eqs.~(\ref{LaNN}) and (\ref{LaND})].
   For the neutral $a_1$ meson we obtain
\begin{equation}
{\cal L}_{a_1 N\Delta}=g_{a_1 N\Delta}\sqrt\frac{2}{3}\begin{pmatrix}\bar{\Delta}^+_\lambda&\bar{\Delta}^0_\lambda\end{pmatrix}
(g^{\lambda\mu}-\gamma^\lambda\gamma^\mu)A_{\mu3}\begin{pmatrix}p\\n\end{pmatrix}+\text{H.c.}
\end{equation}
   In particular, since the external axial-vector field $a_{\mu j}(x)$ and the field $A_{\mu j}(x)$ carry
the same quantum numbers, it is natural to assume the same SU(6) relation for the coupling constants
$g_{a_1N}$ and $g_{a_1 N\Delta}$ as for $\texttt{g}_{A}$ and $\texttt{g}$ [see Eq.~(\ref{relggA})],
\begin{equation}
g_{a_1 N\Delta}=\frac{3}{5}\sqrt{2}g_{a_1N}.
\label{ga1NDga1Nrel}
\end{equation}
   The contribution of the $a_1$ meson to $C_5^A(Q^2)$ is obtained from Eq.~(\ref{CA5lo})
by the replacement
\begin{displaymath}
\texttt{g}\to f_A g_{a_1 N\Delta}\frac{Q^2}{M_{a_1}^2+Q^2}.
\end{displaymath}

   As in the nucleon case, the loop contributions to the low-$Q^2$ behavior of $C_5^A(Q^2)$ are small
\cite{Unal} and we can write
\begin{equation}
C_5^A(Q^2)=g_{AN\Delta}+c_3 Q^2+c_4 \frac{Q^2}{M_{a_1}^2+Q^2},
\end{equation}
where
$c_4=f_A g_{a_1N\Delta}$.
   Extracting $C_5^A(0)=g_{AN\Delta}$, we get
\begin{equation}
C_5^A(Q^2)=g_{AN\Delta}\left[1+\tilde{c}_3Q^2-\tilde{c}_4\frac{(Q^2)^2}{M_{a_1}^2(M^2_{a_1}+Q^2)}\right],
\end{equation}
where $g_{AN\Delta}\tilde{c}_3=c_3+c_4/M^2_{a_1}$ and $\tilde{c}_4=c_4/g_{AN\Delta}$.
   By analogy with Eq.~(\ref{rA2a1}), we find for the mean-square and mean-quartic axial transition radii
\begin{equation}
\langle r^2_{AN\Delta}\rangle=-6\tilde{c}_3,\quad \langle r^4_{AN\Delta}\rangle=-120 \frac{\tilde{c}_4}{M_{a_1}^2}.
\end{equation}

\begin{figure}[h]
\begin{center}
\includegraphics[width=0.8\textwidth]{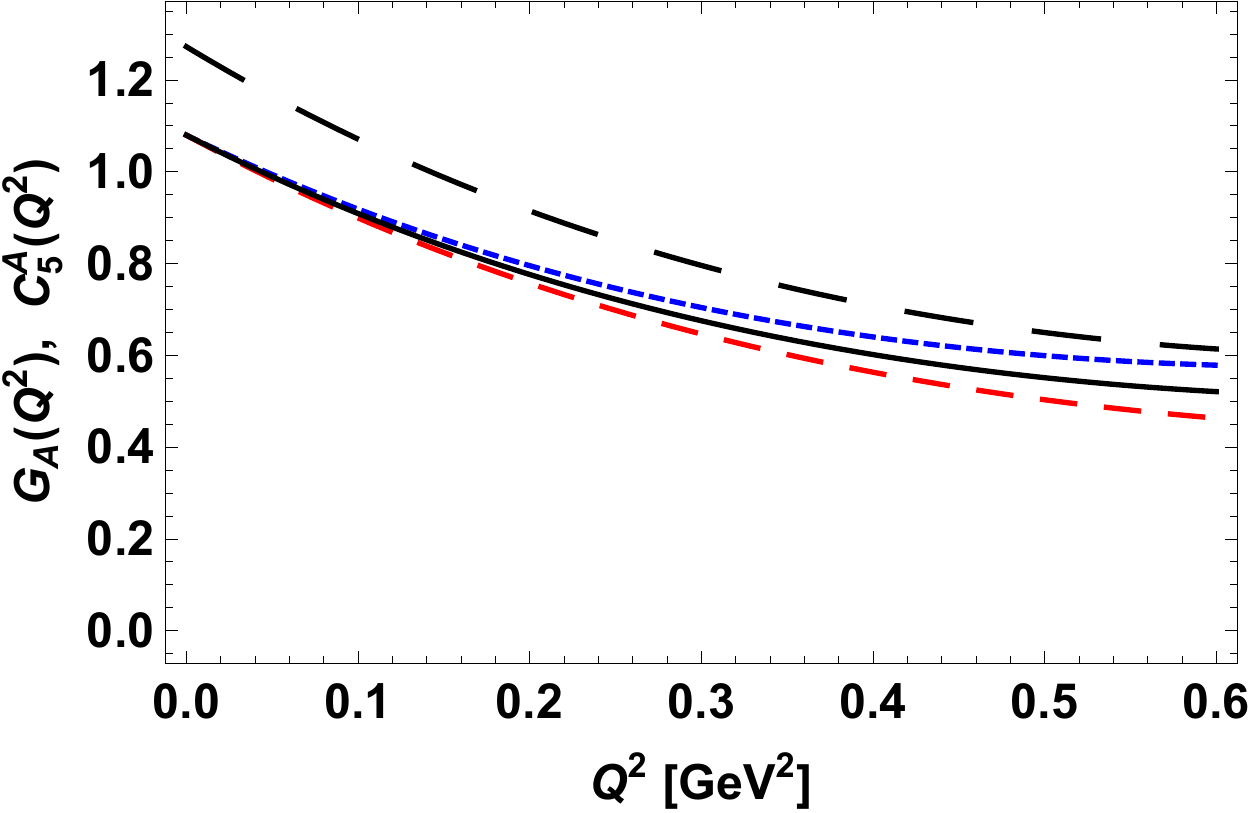}
\caption{\label{figure_GAGAND} (Color online) Axial form factor of the nucleon $G_A(Q^2)$ [(black) long-dashed line].
Axial $N\Delta$ transition form factor $C_5^A(Q^2)$: The (black) solid line corresponds to Eq.~(\ref{C5AGA}),
the (blue) short-dashed and (red) dashed lines correspond to a decrease and an increase of the mean-square
axial transition radius by 5\%, respectively.}
\end{center}
\end{figure}

   At this point, we make use of the quark-model relation of Eq.~(\ref{ga1NDga1Nrel}) between the coupling
constants $g_{a_1N\Delta}$ and $g_{a_1N}$ to reexpress $\tilde{c}_4$ as
\begin{displaymath}
\tilde{c}_4=\frac{f_A g_{a_1N\Delta}}{g_{AN\Delta}}=\frac{3}{5}\sqrt{2}\frac{f_Ag_{a_1N}}{g_{AN\Delta}}
=\frac{3}{5}\sqrt{2}\frac{g_A}{g_{AN\Delta}}\tilde{c}_2.
\end{displaymath}
   Applying, in addition, to $g_{AN\Delta}$ and $g_A$ the quark-model relation of Eq.~(\ref{relggA}), we obtain the simple
result
\begin{equation}
\label{tc2tc4}
\tilde{c}_4=\tilde{c}_2.
\end{equation}
   With these assumptions, the form factor $C_5^A(Q^2)$ contains only one single free parameter $\tilde{c}_3$ (or $c_3$).
   In order to show the dependence on this parameter, as a starting point we make use of the assumption
\begin{equation}
\label{C5AGA}
C_5^A(Q^2)=g_{AN\Delta}F_A(Q^2)=\frac{3}{5}\sqrt{2}\,G_A(Q^2),
\end{equation}
i.e., $\tilde{c}_3=\tilde{c}_1$, and then vary the LEC $\tilde{c}_3$.
   Figure~\ref{figure_GAGAND} shows a comparison between $G_A(Q^2)$
and $C_5^A(Q^2)$.
   The parameters for $G_A(Q^2)$ [(black) long-dashed line] are taken from the fit with
$Q^2_{\rm max}=0.6$~GeV$^2$ (second row of Table \ref{radii_a1}).
   The (black) solid line corresponds to Eq.~(\ref{C5AGA}) for $C_5^A(Q^2)$, the (blue) short-dashed
line and the (red) dashed line correspond to a decrease and an increase of the mean-square
axial transition radius by 5 \%, respectively.

\begin{figure}[h]
\begin{center}
\includegraphics[width=0.8\textwidth]{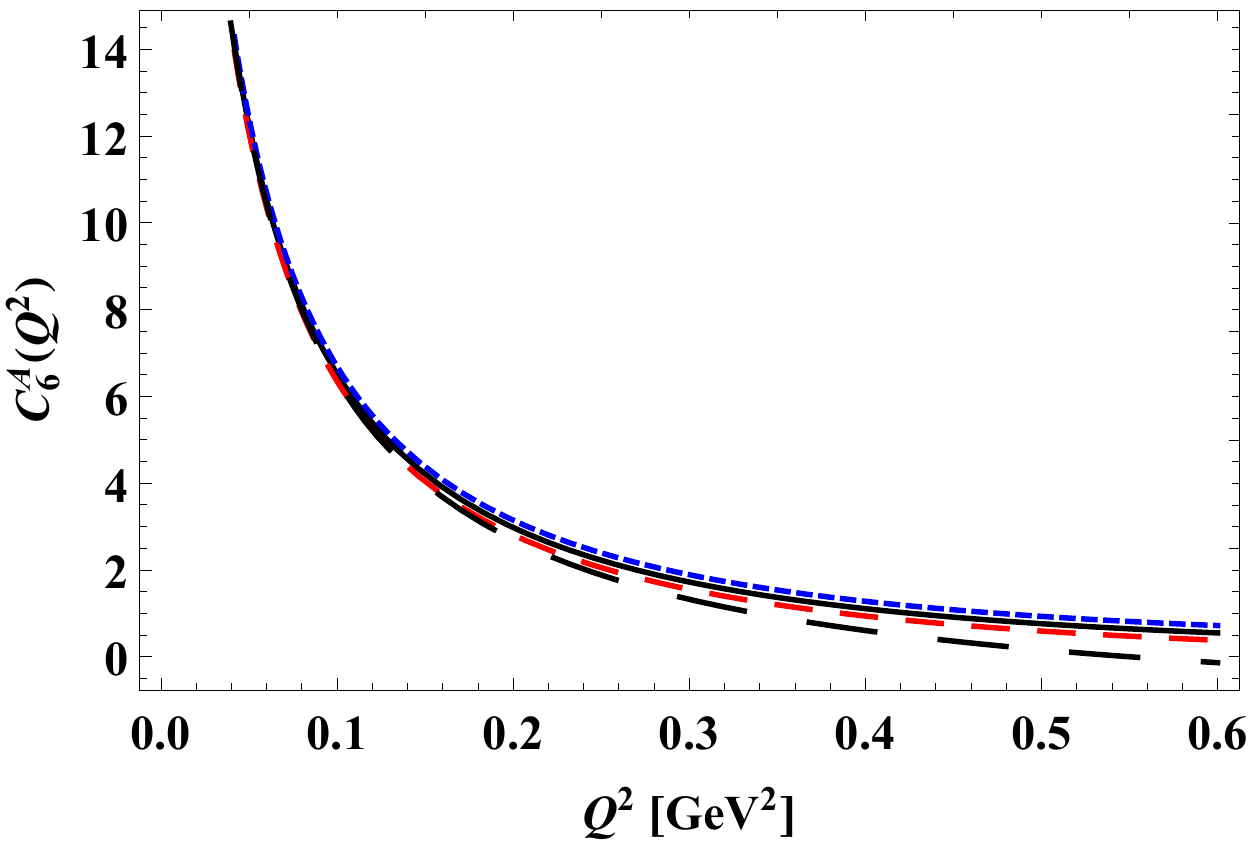}
\caption{\label{figure_CA6} (Color online)
Axial nucleon-to-nucleon transition form factor $C_6^A(Q^2)$.
(Black) long-dashed line: Pion-pole-dominance result of Eq.~(\ref{CA6without}).
The (black) solid line corresponds to Eq.~(\ref{CA6pred}),
the (blue) short-dashed and (red) dashed lines correspond to a decrease and an increase of the mean-square
axial transition radius by 10\%, respectively.}
\end{center}
\end{figure}

   By analogy with Eq.~(\ref{GPexp}), the low-$Q^2$ behavior of $C^A_6(Q^2)$ without the $a_1$ meson can be
written as (see Appendix \ref{appendix_GPiNDelta})
\begin{equation}
\label{CA6without}
C_6^A(Q^2)=\frac{m_N F_\pi g_{\pi N\Delta}}{M_\pi^2+Q^2}+m_N^2 {C_5^A}'(0),
\end{equation}
where $g_{\pi N\Delta}=G_{\pi N\Delta}(-M_\pi^2)$ is the pion-nucleon-$\Delta$ coupling constant.
   Including the $a_1$ meson, we obtain
\begin{equation}
\label{CA6full}
C_6^A(Q^2)=\frac{m_N F_\pi g_{\pi N\Delta}}{M_\pi^2+Q^2}+m_N^2 {C_5^A}'(0)-m_N^2 g_{AN\Delta}\tilde{c}_4\frac{Q^2}{M_{a_1}^2(M_{a_1}^2+Q^2)}.
\end{equation}
   In terms of the lowest-order Lagrangian, Eq.~(\ref{LaND}), and the lowest-order prediction
$g_{AN\Delta}=\texttt{g}$, Eq.~(\ref{CA5lo}), the pion-nucleon-$\Delta$ coupling constant satisfies the generalization of the
Goldberger-Treiman relation \cite{Goldberger:1958tr,Nambu:1960xd},\footnote{Using the values of Table \ref{table_parameters},
the Goldberger-Treiman discrepancy at the nucleon level, $\Delta=1-m_N g_A/(F_\pi g_{\pi N})$, amounts to $\Delta=1.2$\%.}
\begin{equation}
g_{\pi N\Delta}=\frac{m_N}{F_\pi}g_{AN\Delta}.
\end{equation}
   Using $\tilde c_2=\tilde c_4$ of Eq.~(\ref{tc2tc4}) and the quark-model relation $g_{AN\Delta}=3\sqrt{2}g_A/5$, we obtain the following
prediction,
\begin{align}
\label{CA6pred}
C^A_6(Q^2)&=\frac{3\sqrt{2}}{20}G_P(Q^2)+m_N^2\left({C^A_5}'(0)-\frac{3}{5}\sqrt{2}{G_A}'(0)\right)\nonumber\\
&=\frac{3\sqrt{2}}{20}G_P(Q^2)-\frac{1}{6}m_N^2\left(\langle r^2_{AN\Delta}\rangle -\frac{3}{5}\sqrt{2}\langle r^2_A\rangle\right),
\end{align}
where $G_P(Q^2)$ is the induced pseudoscalar form factor of Eq.~(\ref{GPa1}).
   At this stage, we assume that ${G_A}'(0)=c_1$ and ${C^A_5}'(0)=c_3$ are independent.
   Figure~\ref{figure_CA6} shows a comparison between $C_6^A(Q^2)$ without and including the
$a_1$ meson.
   In each case we make use of the quark-model estimate $g_{\pi N\Delta}=3\sqrt{2}g_{\pi N}/5$ as obtained
from the respective Goldberger-Treiman relations.
   The (black) long-dashed line corresponds to Eq.~(\ref{CA6without}) with ${C_5^A}'(0)=g_{AN\Delta}\tilde{c}_1$
and $\tilde{c}_1=-1.78$~GeV$^{-2}$.
   For the result that includes the $a_1$ we assume, in addition, $\tilde{c}_4=\tilde{c}_2=-5.31$ (second row of Table \ref{radii_a1}).
   The (black) solid line corresponds to Eq.~(\ref{CA6full}) for $C_6^A(Q^2)$, the (blue) short-dashed
line and the (red) dashed line correspond to a decrease and an increase of the mean-square
axial transition radius by 10 \%, respectively.

\section{Summary and conclusions}
   We analyzed the low-$Q^2$ behavior of the axial form factor $G_A(Q^2)$, the induced pseudoscalar form factor $G_P(Q^2)$,
and the axial nucleon-to-delta transition form factors $C^A_5(Q^2)$ and $C^A_6(Q^2)$.
   To this end we made use of a chiral effective Lagrangian for the interaction of the $a_1$ meson with
an external axial current, the nucleon, and the $\Delta$.
   Within this approach, the axial form factor $G_A(Q^2)$ is described in terms of three parameters
[see Eq.~(\ref{GAfulla})].
   We investigated the parameters by fitting the model to empirical data, choosing different values
of the maximal squared momentum transfer (see Table~\ref{radii_a1}).
   We compared the results with the commonly used dipole parametrization (see Figs.~\ref{figure_dipole_fit} and \ref{figure_a1_fit}).
   Extending a relation known from chiral perturbation theory, we made a prediction for the induced
pseudoscalar form factor $G_P(Q^2)$.
   For the determination of the transition form factor $C^A_5(Q^2)$ we drew on an SU(6) spin-flavor
quark-model relation to fix $g_{AN\Delta}$ and $g_{a_1N\Delta}$ in terms of $g_A$ and $g_{a_1 N}$, respectively.
   With this assumption, the result for $C^A_5(Q^2)$ depends only on a single parameter $\tilde{c}_3$,
which is related to the mean-square axial transition radius (see Fig.~\ref{figure_GAGAND}).
   Finally, the transition form factor $C^A_6(Q^2)$ was predicted in terms of $G_P(Q^2)$,
and the derivatives ${G_A}'(0)$ and ${C^A_5}'(0)$.
   We emphasize that the predictions at hand represent a model of the
relevant form factors at low $Q^2$.
   To be specific, we expect $Q^2=0.6$~GeV$^2$ to be a reasonable upper limit for the
applicability of the model.

   The purpose of the present investigation was to identify the $a_1$ meson as an important messenger particle
in the context of axial-vector current transitions.
   The use of SU(6) spin-flavor quark-model relations has to be regarded as a first attempt to restrict
the number of free parameters.
   Clearly, merging the $a_1$-meson contribution with the inclusion of pion loops within a consistent power
counting is a desirable next step.
   However, as far as the predictive power is concerned, one has to keep in mind that the chiral effective
field theory calculation will essentially contain the same number of free parameters, i.e., LECs.

\begin{acknowledgments}
Y.~\"U.~and A.~K.~would like to thank the Collaborative Research Center 1044 of the German
Research Foundation for financial support during their stay in Mainz.
The work of Y.~\"U.~was supported by the Scientific and Technological Research Council of Turkey (T\"UB\.{I}TAK).
\end{acknowledgments}

\appendix

\section{Conventions for Dirac spinors}
\label{appendix_Dirac_spinors}
   For the normalization of spinors and states, we follow Appendix A of Ref.~\cite{Gasser:1987rb}.
   We only include the relations which are necessary for our calculation.
\begin{align*}
\langle\vec p\,',r|\vec p, s\rangle&=2E(\vec p)(2\pi)^3\delta^3(\vec p\,'-\vec p)\delta_{rs},\\
E(\vec p)&=\sqrt{m^2+\vec p\,^2},\\
|N(\vec p,s)\rangle&=b^\dagger_s(\vec p)|0\rangle,\\
\{b_r(\vec p\,'),b_s^\dagger(\vec p)\}&=2E(\vec p)(2\pi)^3\delta^3(\vec p\,'-\vec p)\delta_{rs},\\
\Psi(x)&=\sum_{r=1}^2\int\frac{d^3p}{2E(\vec p)(2\pi)^3}\left(b_r(\vec p)u^{(r)}(\vec p)e^{-ip\cdot x}
+d^\dagger_r(\vec p)v^{(r)}(\vec p)e^{ip\cdot x}\right),\\
p^0&=E(\vec p),\\
u^{(r)}(\vec p)&=\sqrt{E(\vec p)+m}\begin{pmatrix}\chi_r\\\frac{\vec\sigma\cdot\vec p}{E(\vec p)+m}\chi_r\end{pmatrix},\\
\chi_1&=\begin{pmatrix}1\\0\end{pmatrix},\quad\chi_2=\begin{pmatrix}0\\1\end{pmatrix},\\
\bar{u}^{(r)}(\vec p)u^{(s)}(\vec p)&=2m\delta_{rs},\\
\langle 0|\Psi(x)|N(\vec p,s)\rangle&=u^{(s)}(\vec p)e^{-ip\cdot x}.
\end{align*}

\section{Low-$Q^2$ expansion of $C_6^A(Q^2)$}
\label{appendix_GPiNDelta}
   We define the pion-nucleon-$\Delta$ form factor $G_{\pi N\Delta}(Q^2)$ in terms of the
reduced matrix element
\begin{equation}
\langle\Delta(p_f,s_f)||\hat m P^{(1)}||N(p_i,s_i)\rangle
=\frac{M_\pi^2 F_\pi}{M_\pi^2+Q^2}G_{\pi N\Delta}(Q^2) i\bar{w}_\lambda(p_f,s_f)\frac{q^\lambda}{m_N}u(p_i,s_i).
\end{equation}
   Using the parametrization of Eq.~(\ref{decomposition1}), the equation for the divergence of
the axial-vector current, Eq.~(\ref{divA}), results in
\begin{equation}
C_5^A(Q^2)-\frac{Q^2}{m_N^2}C_6^A(Q^2)=\frac{M_\pi^2 F_\pi}{M_\pi^2+Q^2}\frac{G_{\pi N\Delta}(Q^2)}{m_N}.
\end{equation}
   Truncating the expansion of the form factors $C_5^A(Q^2)$ and $G_{\pi N\Delta}(Q^2)$ after the linear order in $Q^2$,
\begin{align*}
C_5^A(Q^2)&=C_5^A(0)+Q^2 {C_5^A}'(0),\\
G_{\pi N\Delta}(Q^2)&=G_{\pi N\Delta}(0)+Q^2 {G_{\pi N\Delta}}'(0),
\end{align*}
and using
\begin{displaymath}
g_{\pi N\Delta}=G_{\pi N\Delta}(-M_\pi^2)=G_{\pi N\Delta}(0)-M_\pi^2{G_{\pi N\Delta}}'(0),
\end{displaymath}
we obtain
\begin{align*}
C_6^A(Q^2)&=\frac{m_N^2}{Q^2}\left[C_5^A(Q^2)-\frac{M_\pi^2 F_\pi}{M_\pi^2+Q^2}\frac{G_{\pi N\Delta}(Q^2)}{m_N}\right]\\
&=\frac{m_N^2}{Q^2}\frac{1}{M_\pi^2+Q^2}\left[(M_\pi^2+Q^2)(C_5^A(0)+Q^2 {C^A_5}'(0))
-\frac{M_\pi^2 F_\pi}{m_N}(G_{\pi N\Delta}(0)+Q^2 {G_{\pi N\Delta}}'(0))\right]\\
&=\frac{m_N^2}{Q^2(Q^2+M_\pi^2)}\left[\vphantom{\frac{M_\pi^2 F_\pi}{m_N}}M_\pi^2 C_5^A(0)+M_\pi^2Q^2 {C_5^A}'(0)
+Q^2 C^A_5(0)+(Q^2)^2{C^A_5}'(0)\right.\\
&\quad-\left.\frac{M_\pi^2 F_\pi}{m_N}G_{\pi N\Delta}(0)-\frac{M_\pi^2 F_\pi}{m_N}Q^2 {G_{\pi N\Delta}}'(0)\right]\\
&=\frac{m_N^2}{M_\pi^2+Q^2}\left[C_5^A(0)-M_\pi^2\frac{F_\pi}{m_N}{G_{\pi N\Delta}}'(0)
+(M_\pi^2+Q^2){C_5^A}'(0)\right]\\
&=\frac{m_N^2}{M_\pi^2+Q^2}\left[\frac{F_\pi G_{\pi N\Delta}(0)}{m_N}-M_\pi^2\frac{F_\pi}{m_N}{G_{\pi N\Delta}}'(0)
+(M_\pi^2+Q^2){C_5^A}'(0)\right]\\
&=\frac{m_N F_\pi g_{\pi N\Delta}}{M_\pi^2+Q^2}+m_N^2 {C^A_5}'(0).
\end{align*}

\end{document}